\documentclass[aip,apl,
 amsmath,amssymb,
 reprint,
]{revtex4-1}

\usepackage{graphicx}
\usepackage{dcolumn}
\usepackage{bm}

\usepackage[utf8]{inputenc}
\usepackage[T1]{fontenc}
\usepackage{mathptmx}
\usepackage{etoolbox}
\usepackage{color}
\makeatletter
\def\@email#1#2{%
 \endgroup
 \patchcmd{\titleblock@produce}
  {\frontmatter@RRAPformat}
  {\frontmatter@RRAPformat{\produce@RRAP{*#1\href{mailto:#2}{#2}}}\frontmatter@RRAPformat}
  {}{}
}%
\makeatother
\begin{document}

\preprint{AIP/123-QED}

\title{Ultrafast manipulations of nanoscale skyrmioniums}
\author{H. M. Dong}
\thanks{Authors to whom correspondence should be addressed: hmdong@cumt.edu.cn, yifeng@cumt.edu.cn, and kchang@zju.edu.cn}
\affiliation{School of Materials Science and Physics, China University of Mining and Technology, Xuzhou 221116, China}

\author{P. P. Fu}
\affiliation{School of Materials Science and Physics, China University of Mining and Technology, Xuzhou 221116, China}

\author{Y. F. Duan}
\thanks{Authors to whom correspondence should be addressed: hmdong@cumt.edu.cn, yifeng@cumt.edu.cn, and kchang@zju.edu.cn}
\affiliation{School of Materials Science and Physics, China University of Mining and Technology, Xuzhou 221116, China}

\author{K. Chang}
\thanks{Authors to whom correspondence should be addressed: hmdong@cumt.edu.cn, yifeng@cumt.edu.cn, and kchang@zju.edu.cn}
\affiliation{School of Physics, Zhejiang University, Hangzhou 310027, P. R. China}

\date{\today}
\begin{abstract}
The advancement of next-generation magnetic devices depends on fast manipulating magnetic microstructures on the nanoscale. A universal method is presented for rapidly and reliably generating, controlling, and driving nano-scale skyrmioniums, through high-throughput micromagnetic simulations. Ultrafast switches are realized between skyrmionium and skyrmion states and rapidly change their polarities in monolayer magnetic nanodiscs by perpendicular magnetic fields. The transition mechanism by alternating magnetic fields differs from that under steady magnetic fields. New skyrmionic textures, such as flower-like and windmill-like skyrmions, are discovered. Moreover, this nanoscale skyrmionium can move rapidly and stably in nanoribbons using weaker spin-polarized currents. Explicit discussions are held regarding the physical mechanisms involved in ultrafast manipulations of skyrmioniums. This work provides further physical insight into the manipulation and applications of topological skyrmionic structures for developing low-power consumption and nanostorage devices.
\end{abstract}

\maketitle
\section{Introduction}
Topologically protected magnetic skyrmions have attracted much attention since their experimental discovery, especially with the promise of being used as magnetic memory devices, such as racetrack memories. However, the spin-polarized current-driven magnetic skyrmion appears to move with the transverse velocity due to the skyrmion Hall effect (SkHE), which affects its stability and greatly limits its applications. The ultrafast manipulation of magnetic topologies at the nanoscale remains a significant challenge. Moreover, new topological spin textures, such as skyrmioniums, merons, and biskyrmions, have been discovered and studied \cite{gobel_beyond_2021, yang2023}, which opens up possibilities for exploring new magnetic topological properties and addressing the limitations of skyrmions.

A skyrmionium, namely $2\pi$-skyrmion with a zero topological charge $(Q=0)$, can be regarded as the combination of two magnetic skyrmions with opposing topological charges $(Q = \pm 1)$, which have been experimentally observed in ferromagnet-magnetic topological insulator heterostructures \cite{zhang_2018}. The skyrmioniums driven by a spin-polarized current or a spin wave can move without a Hall angle or SkHE, and as a result, have been attracting increasing attention recently \cite{yang2023,shen_2018}. An earlier study found that a stable static skyrmionium is generated in a nanodisk with a relaxed time of up to 0.4 ns \cite{control_2016}. It is found that skyrmioniums can be formed in 2D Janus magnets due to the competition between Dzyaloshinskii-Moriya interaction (DMI) and exchange frustration \cite{zhangy_2020}. Skyrmions and skyrmioniums are stabilized in confined ultrathin ferromagnetic (FM) nanodisks, originating from the competition between exchange interactions and magnetic anisotropies \cite{bo2020}. The stale skyrmionium is presented and transitions between skyrmions and skyrmioniums are achieved, owing to the geometrical confinement of magnetic hemispherical shells \cite{yang_2021}. The antiferromagnetic skyrmionium can be generated by a local injection of spin current and propagate under spatially uniform spin currents \cite{obadero_2020}. Very recently, it is found that a skyrmionium can be transited to a skyrmion through the thermal annihilation of the inner skyrmion \cite{PR013229}. Robust-driven nanoscale skyrmioniums and ultrafast switching, however, still require further investigation so far. 

Furthermore, nanoscale Bloch-type skyrmioniums can be stabilized and directionally driven by the damping-like spin-orbit torque (SOT) in a monolayer frustrated magnet\cite{xia_2020}. In addition, it has been demonstrated that skyrmioniums driven by SOT are more resistant to distortion in ferrimagnets due to the zero intrinsic SkHE \cite{liang_2021}. Very recently, the reversible conversion between micron-scale skyrmions and skyrmioniums has been experimentally realized in a ferromagnetic multilayer system by a sinusoidal current pulse under a perpendicular magnetic field \cite{yang2023}. This was achieved by applying a sinusoidal current pulse under a perpendicular magnetic field, showcasing the versatility of these topological spin structures and their potential for further development. However, the fast manipulation and driving of stable nanoscale skyrmioniums by a simple and efficient method in FM systems have been a major obstacle in the magnetism of science and technology at present, particularly for driving by a spin transfer torque (STT) \cite{masell214428}. This challenge persists due to the inherent complexities and limitations associated with the dynamics and stability of these nanoscale magnetic configurations in FM systems.

Although skyrmioniums have been extensively studied, fast manipulation of skyrmioniums recently, as well as transitions to and from the other non-trivial topological states, is still very limited \cite{bo2020,kolesnikov_2018,ponsudana_2022,VIGOCO166848,PR013229,mehmood2020}. The skyrmionium nucleation and motion in a Pt/Co/Ta nanotrack with the enhancement of the spin Hall effect have been confirmed \cite{kolesnikov_2018}. The switching from a single domain to skyrmion or skyrmionium can be realized by an oscillating perpendicular magnetic field \cite{VIGOCO166848}. The nucleation of skyrmionium on a nanodisk has been demonstrated by applying an oscillating perpendicular magnetic field with a frequency equaling the eigenfrequency of skyrmionium in a nanoring \cite{ponsudana_2022}. The skyrmion and skyrmionium are stabilized well due to the competition between exchange interactions and anisotropy contributions. The transition between the skyrmion and the skyrmionium can be achieved through fine-tuning the magnetic anisotropy \cite{bo2020}. Furthermore, it is found that the strain-driven switching of magnetic skyrmion and other topological states in a nanodisc of trilayer Pt/Co/Ta stacks can be possible \cite{mehmood2020}. Addressing this challenge is of paramount importance, as the ability to reliably and efficiently manipulate and drive nanoscale skyrmioniums could unlock a wide range of innovative applications in areas such as high-density data storage, logic devices, and neuromorphic computing. Continued research and development in this field are crucial for overcoming the current limitations and propelling the field of magnetism science and technology forward.

In this letter, we achieve stable nano-skyrmioniums with about 18 nm diameters in a monolayer FM nanodisk with perpendicular magnetic anisotropy (PMA). Such nanoscale skyrmioniums driven by the STT of weaker spin-polarized currents, can move stably and rapidly in nanoribbons. In particular, an ultrafast switching between skyrmioniums and skyrmions, as well as rapid shifts in topological charge polarity, can be realized using magnetic fields. New-type skyrmionic textures with nonzero topological charges are observed, which implies that there may be more novel non-trivial topological magnetic structures even in a simple FM monolayer. Note that the outcomes from the recent experiment resemble the conclusions drawn from our theoretical discoveries \cite{yoshimochi_2024}. The proposed method is simple, practical, and experimentally easy to implement. This further contributes to the potential of skyrmionium-based nanomemories for the next generation.

\section{Results and Discsussions}
\subsection{Phase diagrams}
By utilizing high-throughput micromagnetic simulations, we first obtain the detailed phase diagrams for the exchange interaction $A$, the interface-induced DMI constant $D$, and the PMA constant $K_u$ that demonstrate the 2D magnetic structures, as shown in Fig. 1 (a), (b) and (c). The competition between the exchange interaction and the PMA, which favors collinear magnetizations, and intrinsic DMI, which prefers noncollinear chiral magnetizations, leads to the emergence of new and unique spin textures. The detailed micromagnetic simulation method and descriptions can be found in the supplementary materials (SMs). Skyrmion states are produced when the values of these parameters are roughly equal for the three parameters $A$, $D$, and $K_u$. These phase diagrams indicate that the conditions for forming skyrmions and skyrmioniums are very similar, while skyrmioniums require higher $D$ and lower $K_u$. The formation of skyrmioniums is more difficult than skyrmions since the energy of skyrmioniums is slightly higher than that of skyrmions. An observation made in Phase III is that the presence of skyrmions and skyrmioniums can be noted in Fig. 1 (d), and kindly refer to Fig. S1 located in the SMs. 
\begin{figure}[!t]
\centering{}\includegraphics[width=1.0\columnwidth]{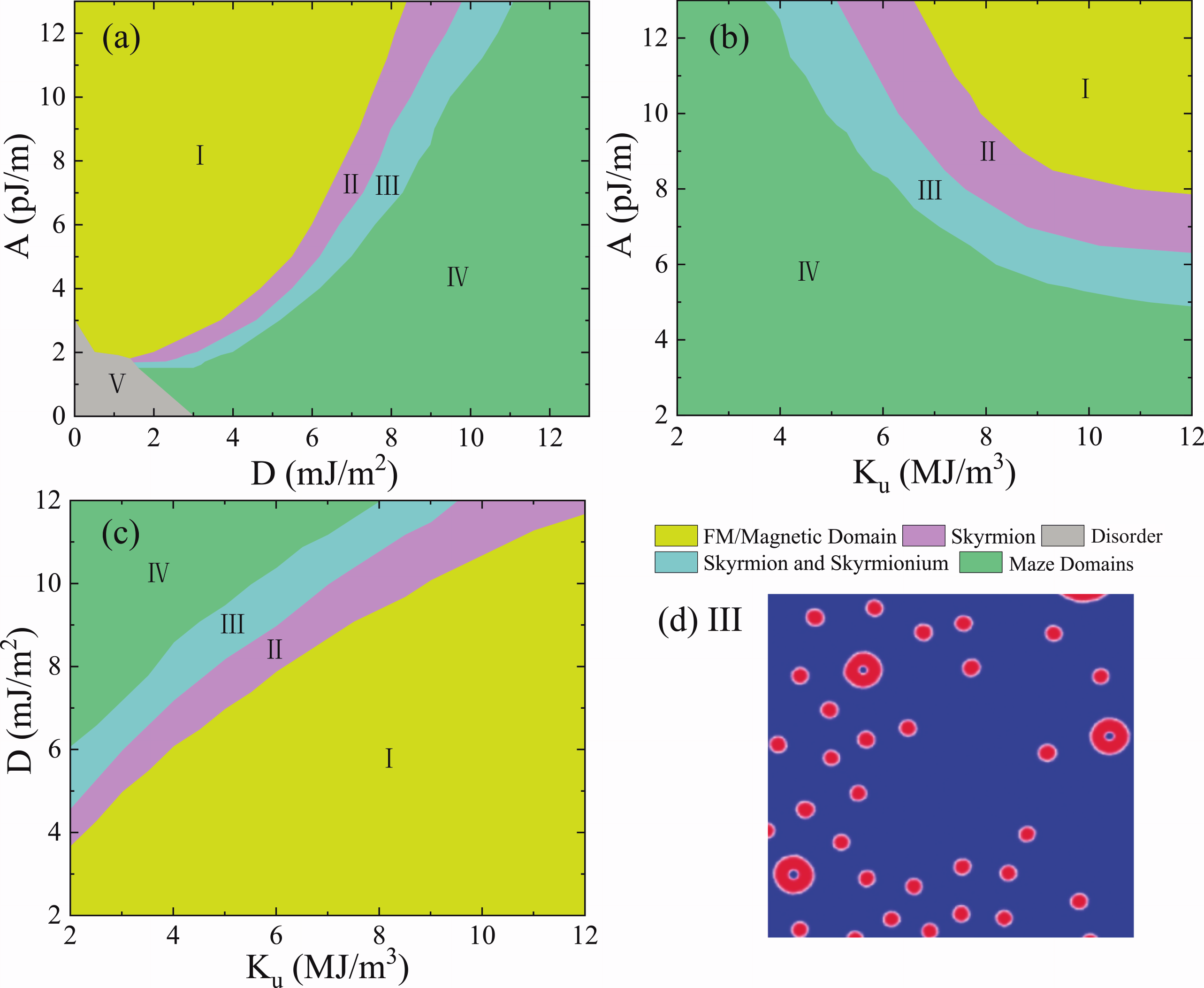}
\caption{(a), (b) and (c) Phase diagram for the magnetic structure of a single-layer magnet, (d) Skyrmions and skyrmioniums in phase III.}
\label{fig1}
\end{figure}

Our study shows that skyrmions are formed for $\kappa = \pi^2D^2/(16AK_u) < 1$, while skyrmions and skyrmioniums can coexist, i.e., forming phase III when $0.5 < \kappa < 1$ \cite{dong2023}. Skyrmions coexist together in both positive and negative topological charge polarities $(Q=\pm1)$, which offers the possibility of realizing a switch in the polarity of the topological charges. The formation of skyrmioniums requires the conditions $0.5 < \kappa < 1$. Skyrmion bags and skyrmioniums are remarkably alike in their formation, which is in agreement with the recent experimental results \cite{yang2023}. The maze domain states are generated when $D$ is strong enough, as shown in Fig. 1. The skyrmioniums in phase III split into skyrmions as $A$ increases, while they evolve into maze states for the increasing of $D$. Furthermore, these phase diagrams can be used to screen materials to realize the appropriate topological magnetic structures experimentally. Our simulations show that skyrmionium states are metastable in the presence of free boundaries, and thus require the introduction of a new mechanism to be stabilized.  For free boundaries, there are no restrictions imposed on the magnetizations and the magnetizations are free to align and orient in any direction at the boundaries \cite{rybakov_new_2015}. Certain restrictions to the order parameters imposed at the boundary can lead to the formation of new topological states \cite{rybakov_new_2015}.

\subsection{Driving Skyrmioniums}
\begin{figure}[!t]
\centering{}\includegraphics[width=1.0\columnwidth]{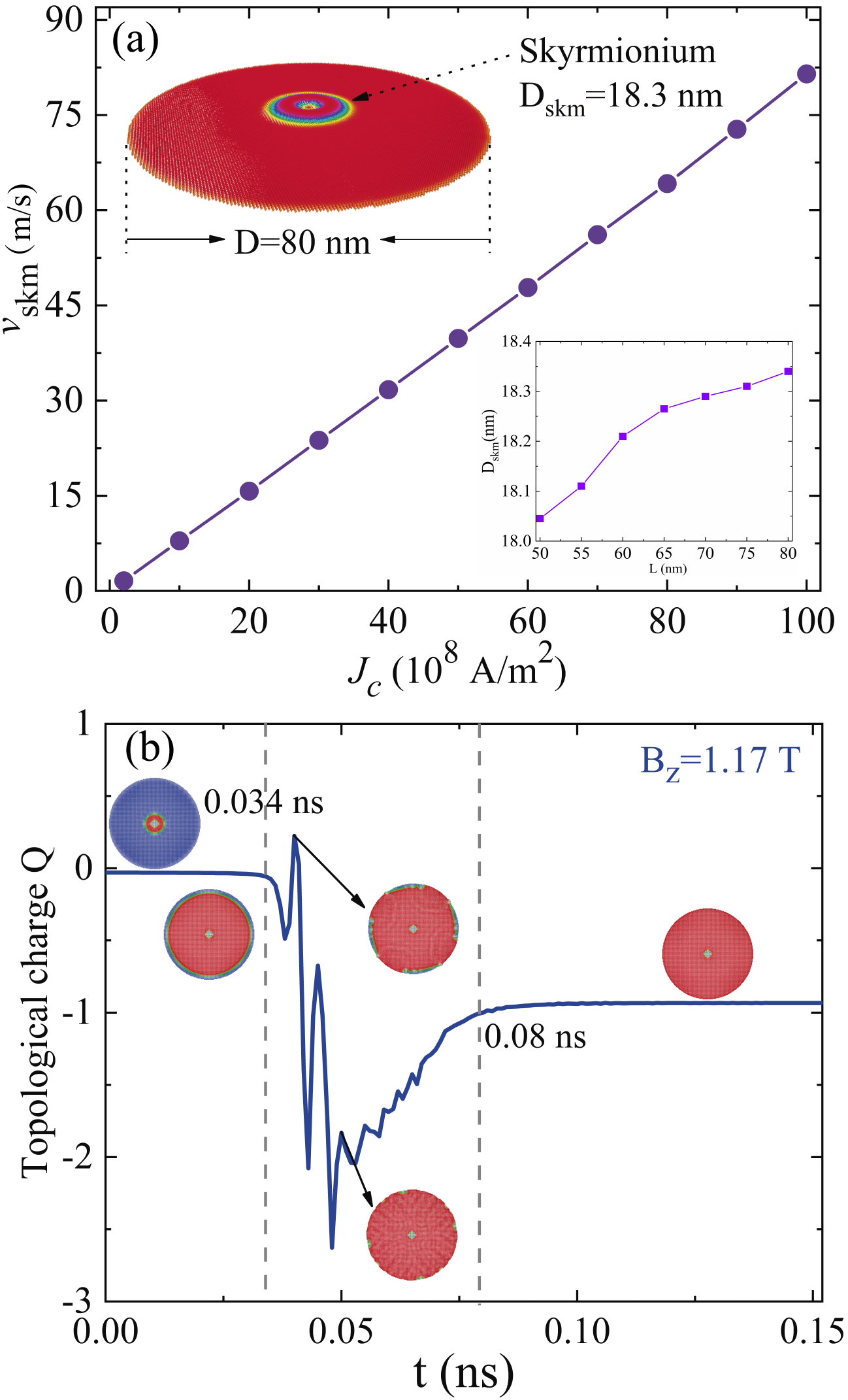}
\caption{(a) The velocities $v_{skm}$ of skyrmioniums as the function of the polarization current density $J_c$. The diameters $D_{skm}$ and the out-of-plane magnetic configuration $m_z$ of skyrmioniums depend on the diameters $L$ of FM nanodisks (insert). (b) The temporal evolution of the topological charge and the transition processes of the corresponding magnetization profiles in the presence of a steady perpendicular magnetic field.}
\label{fig2}
\end{figure}

In FM nanostructures, the edges play a major role and the edge magnetization rotates in a plane \cite{rohart_2013,quispe_2020}. As a result, isolated single noncollinear magnetic textures can be formed in such nanostructures, with the period $L_0=4\pi A/D$ of noncollinear spiral states \cite{rohart_2013}. Based on these findings, we propose that FM nanodiscs with PMA and $L_0=15.6$ nm (see S1 in SMs) can stabilize N$\acute{e}$el-type skyrmioniums by massive magnetic simulations, resulting from the edge effect due to symmetry-breaking and geometrical confinement effect \cite{charilaou_2018}. Therefore, the diameter $L$ of the nanodisks that stabilize skyrmions is around $L=2L_0=31.3$ nm, while $L$ for supporting skyrmioniums is around $L=4L_0=62.6$ nm.  Due to the PMA effect on the spiral period, a range of diameters is possible in nanostructures \cite{rohart_2013}. Our study demonstrates that magnetic disks with the diameters $L=50-80$ nm can yield stable isolated skyrmioniums with the diameter approximate $D_{skm}\sim L_0=18.2$ nm. For $L < 50$ nm, the skyrmionium collapses into a single skyrmion, and for $L > 80$ nm, no individual skyrmionium can be produced. If the diameter $L$ continues to increase, a distinctive $3\pi$-skyrmion with the topological charge $Q=-1$ forms in the center (see Fig. S2 in SMs), for example, $L=6L_0=93.8$ nm. Such topologically nontrivial textures can coexist with skyrmion states, and their spin rotates $3\pi$ from the center to the edge. Notably, the diameters of these skyrmioniums exhibit minimal variation with nanodisk sizes $L=50-80$ nm in the insert of Fig. 2 (a). As a result, the method yields skyrmioniums that possess remarkable robustness in their physical sizes. We present a stable nano-skyrmionium with a diameter $D_{skm}\simeq 18.3$ nm generated in such FM nano-disks with a diameter of $L=80$ nm, which can be driven by the STT induced by low spin-polarized currents in a nano-racetrack memory, as shown in Fig. 2 (a). 

The stale terminal velocity $v_{skm}$ of the skyrmionium shows a very perfect linear relationship with the low polarization current density $J_c$ in Fig. 2 (a). Please refer to Supplementary Video 1 for details. This typical linear response relationship $v_{skm}\sim J_c$ is due to its stable topological structure and the absence of movement in the perpendicular current direction for low driving current density $J_c$. The velocity is up to 80 m/s without SkHE when driven by an electric current $J_c=1 \times 10^{10}$ A/m$^2$. In comparison to the current density driving magnetic domain walls or skyrmions, the spin current density is much lower. The linear dependence on the current density of FM skyrmioniums is very similar to that of antiferromagnetic skyrmioniums \cite{obadero_2020}. With a measurement of 18.3 nm in diameter (insert in Fig. (2)), the size is considerably smaller than that of a typical skyrmion in magnetic materials. Please also see Figs. S2 in the SMs. The skyrmionium can move quickly and consistently without deformations when $J_c < 1.3\times 10^{10}$ A/m$^2$. Our achieved skyrmioniums are indeed remarkable, for they remain non-distortion in both nano-sizes and shapes as they move (see Figs. S3 and S4 in the SMs), surpassing the previous experimentations \cite{yang2023,zhang_2018}. Very recently, it has been experimentally found that the diameter of the skyrmionium decreases when moving along the direction of the current \cite{yang2023}. In this work, we find that the proposed nanoscale skyrmioniums without distortion can move fast and solidly under weak spin current densities, which is exactly needed to fabricate the next generation of low-power and high-density memory devices. 

\subsection{Transition by steady magnetic fields}
The ability to quickly switch between different magnetic topological states is very beneficial for the manipulation of spin structures to develop next-generation magnetic memory devices. The topological charge (number) $Q$ of the system is calculated as $Q=\frac{1}{4\pi}\iint n_{{\mathrm{sk}}}\mathrm{d}x\mathrm{d}y$ with the 2D topological charge density $n_{\mathrm{sk~}}=\mathbf{m}\cdot(\partial_x\mathbf{m~}\times\partial_y\mathbf{m})$ \cite{dong2023}. In this work, we show that such a skyrmionium can transmit into a stable skyrmion ($Q=-1$) within 0.1 ns when a steady magnetic field B $=1.17-2.92$ T is applied in the perpendicular direction, such as $B=1.17$ T in Fig. 2 (b). Superconducting magnets are remarkable devices that can generate exceptionally powerful magnetic fields. Superconducting magnets can experimentally produce extremely strong magnetic fields with the strength of 8-9 Tesla when cooled to extremely low temperatures\cite{Dong21}. The strength B of the steady magnetic field that induces the transition depends both on the diameter $L$ of the nanodisc and the polarity of skyrmioniums, i.e., the direction of the central spin. However, this skyrmion will not turn into a skyrmionium if the magnetic field is turned off (i.e. $B=0$) at this point after a sufficiently long relaxation time. This is because, in comparison to skyrmions, skyrmioniums are metastable states with higher energies as shown in Fig. S5 (a). After the transformation, the nanosystem mostly turns into collinear magnetizations. The magnetic exchange energy (including DMI energy) increases. However, the magnetic field causes a considerable reduction in the magnetic anisotropy energy, leading to a decrease in the total magnetic energy of the whole system [see Fig. S5 (c)]. 

The stability of skyrmioniums is weaker than that of skyrmions. This indicates that to achieve the quick inversion of skyrmions into skyrmioniums, new mechanisms, like external fields, may need to be introduced. Additionally, we also find that the switching we achieved is much faster than that of the skyrmioniums discovered recently \cite{yang_2021,obadero_2020}. It is noticed that the topological charge must be an integer. As a result, we find that when non-trivial topological states transition into each other, the transition is generally discontinuous. In contrast, when trivial topological configurations transition into each other, the transition is generally continuous.

\subsection{Switching by alternating magnetic fields}
\begin{figure}[!t]
\centering{}\includegraphics[width=1.0\columnwidth]{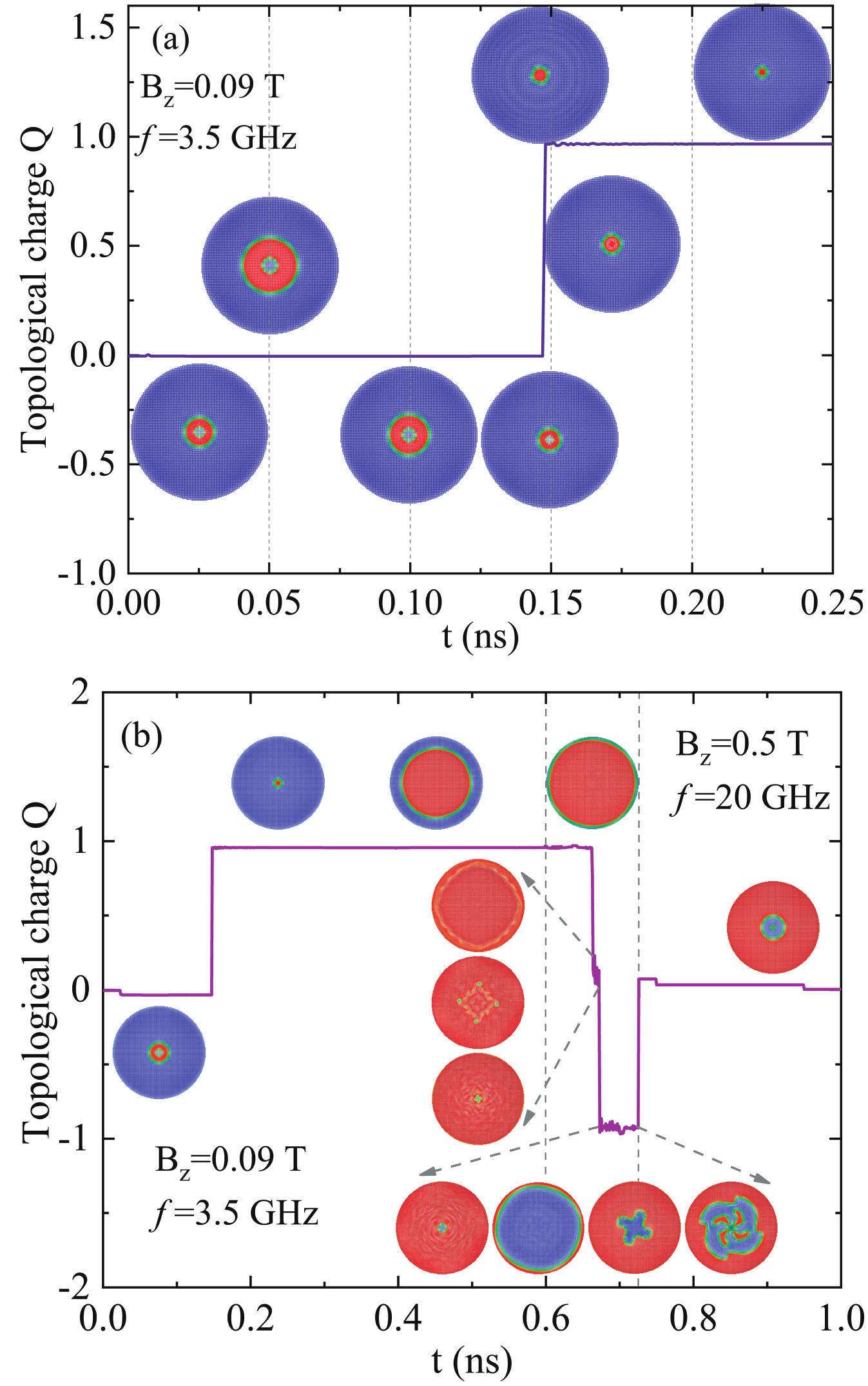}
\caption{The temporal evolution of the topological charge and the transition processes of magnetic configurations under a perpendicular alternating magnetic field: (a) transmission from a skyrmionium to a skyrmion, and (b) switching between a skyrmionium and a skyrmion.}
\label{fig3}
\end{figure}

The alternating sinusoidal magnetic field $B=B_z\sin(2\pi ft)$ with the frequency $f$ along the z-axis is applied for skyrmioniums in $L=80$ nm nanostructures. We find that the previous skyrmionium transforms into a skyrmion with $Q=+1$ for $B_z=0.09$ T and $f=3.5$ GHz within 0.2 ns in Fig. 3 (a), in contrast to applying a steady magnetic field. Please refer to Supplementary Video 2 for further details. The frequency $f$ can be determined by the spin wave modes of skyrmions or skyrmioniums with the power spectrum of the magnetization $m_z$ \cite{zhao_direct_2016,yang_2021}. Here $f=3.5$ GHz is the spin wave modes of skyrmions in the nanodiscs. Consequently, for the skyrmion without the alternating magnetic field ($B=0$), the skyrmion does not revert to the original skyrmionium, similar to a steady magnetic field. Interestingly, we find that when a perpendicularly oscillating magnetic field with $B_z= 0.5$ T and $f = 20$ GHz is applied to the stabilized skyrmion state, it transforms back into a stabilized skyrmionium. It is interesting to note that when the stable skyrmion state is subjected to a perpendicularly oscillating magnetic field with $B_0= 0.5$ T and $f=20$ GHz, the skyrmion with $Q = +1$ will first transform into a skyrmion state with $Q = -1$. Here $f=20$ GHz is the spin wave modes of skyrmioniums. Finally, it converts into a stable skyrmionium with an opposing polarity, as shown in Fig. 3 (b). This indicates that an alternating magnetic field can alter both the magnetic topological state and the corresponding polarity in FM nano-structures. The transition mechanism under an alternating magnetic field differs from that under a steady magnetic field. While the latter is a competition and redistribution of micromagnetic energy due to steady magnetic fields, the former induces spin-flips via spin-wave modes resulting from alternating magnetic fields. Therefore, we find that the change in micromagnetic energies during the transition is negligible under an alternating magnetic field, as shown in Fig. S5 (b) and (d) of SMs. The strong magnetic field at GHz frequency is not easy to achieve in experiments, and understanding the high-frequency magnetic properties of materials at GHz frequency is challenging. However, there has been recent discussion about commercially available experimental platforms with components that operate at frequencies up to 60 GHz \cite{drost8137, jiang2015}.

Moreover, some novel non-trivial topological states are observed, such as flower-like and windmill-like skyrmions. We show the magnetization configurations of the skyrmionium, the flower-like and the windmill-like skyrmions in detail, and the corresponding simulated magnetic force microscopy images in Fig. (4). Our study insightfully shows that some dynamic skyrmion states with the topological charge $Q=\pm1$ can be created in confined magnetic nano-systems. Despite their simplicity, even single-layer magnetic nanostructures support complex and non-trivial topological textures, as shown in Fig. S6 of SMs. It is noticed that directly imaging the evolution of highly geometrically confined individual skyrmions and their magnetic field-driven transitions has been realized in FeGe nanodisks by using Lorentz transmission electron microscopy \cite{zhao_direct_2016}. These non-trivial magnetic topological states have not been discovered before. We hope these novel topological states will be further investigated theoretically and experimentally in future research. These non-trivial topological magnetic structures are dynamic in alternating magnetic fields. New mechanisms are needed to form such static and stable magnetic structures, such as size and shape confinement effects in nanostructures. We are exploring the possibility of realizing these new, non-trivial topological in nano-heterostructures with different shapes.

The intricate and captivating magnetic structures observed in materials, such as the flower- and windmill-like skyrmions, exhibit a remarkable dynamism under the influence of an alternating magnetic field. These magnetic configurations are not merely static but rather evolve and transform periodically, synchronizing with the oscillating nature of the applied magnetic field. This dynamic behavior is a testament to the complex interplay between the material's internal magnetic interactions and the external magnetic environment in magnetic nanostructures, which should be studied in depth \cite{noceri2023}.

\begin{figure}[!t]
\centering{}\includegraphics[width=1.0\columnwidth]{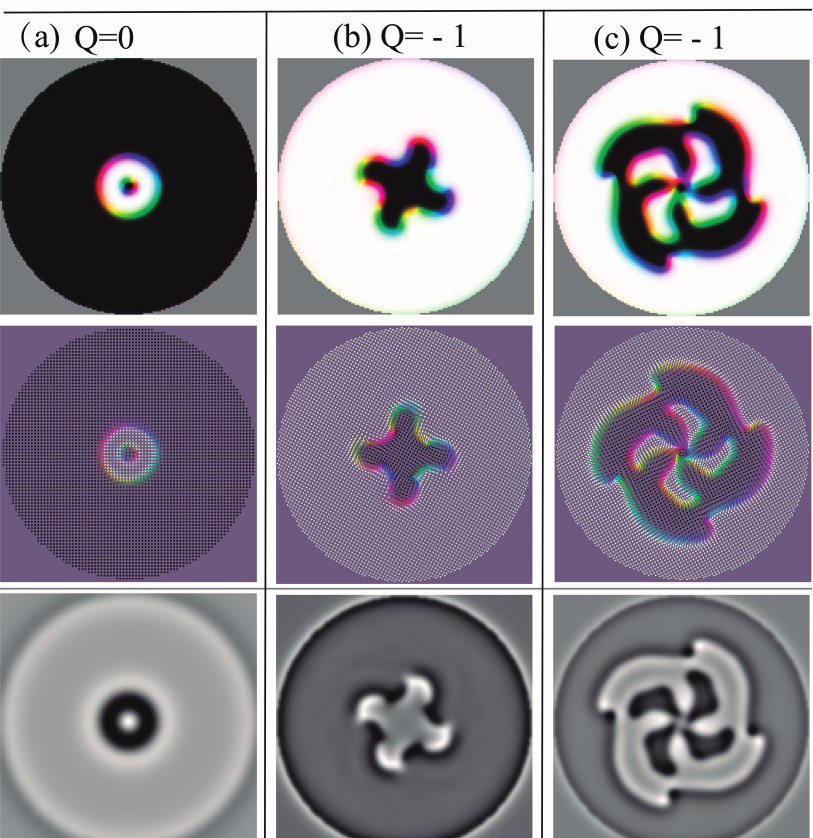}
\caption{The magnetization configurations of the skyrmionium (left), the flower-like (center) and the windmill-like (right) skyrmions, and the corresponding simulated magnetic force microscopy images (bottoms).}
\label{fig4}
\end{figure}
The dynamic nature of these magnetic structures presents both challenges and opportunities for researchers and engineers. To fully harness the potential of these fascinating magnetic phenomena, new mechanisms and approaches must be developed to reliably create and maintain static and stable magnetic structures \cite{kim_2020}. This pursuit will require a deep understanding of the underlying physical processes, as well as the development of innovative material design strategies and experimental techniques. The exploration of these dynamic and topological magnetic structures holds the promise of unlocking new frontiers in areas such as data storage, magnetic sensing, and energy-efficient computing. Presently, there is currently very little research on this scientific topic at this time. We hope to delve deeper into this captivating realm of magnetism, and the potential for groundbreaking discoveries and technological advancements.

\section{Conclusion}
In conclusion, stable ultrafast-moving nano-skyrmioniums with about 18 nm diameter are achieved without the SkHE, driven by relatively weak spin-polarized currents, owing to the edge effect and the geometrical confinement effect. The quick conversion of skyrmioniums to skyrmions can be accomplished by a steady magnetic field. We realize a fast switch between skyrmioniums and skyrmions via an alternating magnetic field, considering the the spin wave modes of magnetic textures. Novel non-trivial topological structures, such as $3\pi$-skyrmions, flower-like and windmill-like skyrmions, have been discovered which require further investigation. Our investigation expands the potential for controlling the shape and structure of topological spin textures, leading to the creation of approaches for manipulating spin textures.

\vspace{12 pt}
\section*{Supplementary Material}
See the supplementary material for the micromagnetic simulations, the magnetization configurations, the dynamics of skyrmioniums, the temporal evolution of magnetic energy, the magnetic force microscopy images, and the spin textures.

\begin{acknowledgments}
This work is supported by the Key Academic Discipline Project of China University of Mining and Technology (No. 2022WLXK03), by the National Natural Science Foundation of China (Grant No. 12374079).
\end{acknowledgments}
\section*{AUTHOR DECLARATIONS}
{\bf Conflict of Interest}\par
The authors have no conflicts to disclose.
\section*{DATA AVAILABILITY}
The data that support the findings of this study are available from the corresponding authors upon reasonable request.

\nocite{*}
\providecommand{\noopsort}[1]{}\providecommand{\singleletter}[1]{#1}%

\end{document}